\documentclass[journal]{IEEEtran}

\usepackage{graphicx}
\usepackage{url}
\usepackage[hide]{ed}
\usepackage{color, soul}
\usepackage{caption}

\begin{document}

\title{Process Monitoring on Sequences\\ of System Call Count Vectors}

\author{Michael Dymshits, Benjamin Myara, David
Tolpin\thanks{David Tolpin was with PayPal at the time of
preparing the paper for publication.}\\PayPal\\mdymshits@paypal.com, bibi@paypal.com,
dvd@offtopia.net}

\IEEEpubid{978-1-5386-1585-0/17/\$31.00 \copyright 2017 IEEE}

\maketitle

\begin{abstract}
    We introduce a methodology for efficient monitoring of processes
    running on hosts in a corporate network. The methodology is
    based on collecting streams of system calls produced by all
    or selected processes on the hosts, and sending them over
    the network to a monitoring  server, where machine learning
    algorithms are used to identify changes in process behavior
    due to malicious activity, hardware failures, or software
    errors. The methodology uses a sequence of system call
    count vectors as the data format which can handle
    large and varying volumes of data.

    Unlike previous approaches, the methodology introduced in
    this paper is suitable for distributed collection and
    processing of data in large corporate networks. We evaluate
    the methodology both in a laboratory setting on a real-life
    setup and provide statistics characterizing performance and
accuracy of the methodology.
\end{abstract}


\section{Introduction}

System call streams are enormous, and an efficient
representation with performance guarantees independent of the
level of activity on the host must be used.  Some earlier work
was based on processing of sequential streams of system
calls~\cite{FHS96,WFP99}, which does not scale well --- a single
process can produce tens of thousands system calls per second,
with hundreds of processes running on each host, or \textit{end
point}. Other approaches rely on computing frequencies of short
sequences (n-grams) of system calls over a fixed time
window~~\cite{LMH+05,WSA+13}. However, in this case information
about temporal dynamics of the process is lost.

Further on, both from security and performance points of view
some of the processing is sent from the monitored host to the
\textit{monitoring server} --- a different machine, dedicated to
the monitoring task. This poses additional restrictions on the
amount of data which can be collected: on the one hand, the network
load must stay within the allowed limits; on the other hand, the
machine executing the monitoring task must be able to process
data from multiple hosts in the network.

In this paper we introduce a new methodology for monitoring
networked computer systems based on system calls. The
methodology combines careful selection of information being
gathered with employment of advanced machine learning
algorithms. We evaluate the methodology on a reproducible 
real-life setup, as well as provide statistics for
production-level deployment of a monitoring system based on the
methodology.

The paper proceeds as follows. Section~\ref{sec:related} surveys
related work on system call based monitoring.
Section~\ref{sec:overview} describes the overall structure of
the solution and summarizes results of the empirical evaluation.
Section~\ref{sec:details} provides detailed explanation and
justification of the solution architecture and technological
choices, as well as addresses issues of data collection.
Section~\ref{sec:empirical} provides empirical evaluation of the
methodology on a real-life setup, as well as statistics of a
production-level deployment. Finally, Section~\ref{sec:summary}
summarizes paper contributions and suggests directions for future
work.

\IEEEpubidadjcol

\section{Related Work}
\label{sec:related}

Research of system-call based techniques for process
identification and anomaly detection has been conducted since
the 1990s.  \cite{FHS96} is the seminal work which pushed
forward research on methods and representations of operating
system process monitoring based on system call. 

Main research directions are methods and models of process
behavior, on the one hand, and representation of system calls and
system call sequences, on the other hand.

\cite{WFP99} provides an early comparison of machine learning
methods for modeling process behavior. \cite{GRS04} introduces
the model of execution graph, and behavior similarity measure
based on the execution graph. \cite{MVV+06} combines multiple
models into an ensemble to improve anomaly detection.
\cite{XS10} applies continuous time Bayesian network (CTBN) to
system call processes to account for time-dependent features and
address high variability of system call streams over time.
\cite{KYL+16} applies a deep LSTM-based architecture to
sequences of individual system calls, treating system calls as a
language model.

Initially, only system call indices were used as
features~\cite{FHS96,WFP99}. \cite{LMH+05} compares three
different representations of system calls: n-grams of system call
names, histograms of system call names, and individual system
calls with associated parameters. \cite{PS07} proposes the use of
system call sequences of varying length as features.
\cite{LMH+05,TC06} investigate extracting features for machine
learning from arguments of system calls.  \cite{WSA+13} studies
novel techniques of anomaly detection and classification using
n-grams of system calls. \cite{CMK15} conducts an case study of
n-gram based feature selection for system-call based monitoring,
and analyses the influence of the size of the n-gram set and the
maximum n-gram length  on detection accuracy. 

This work differs from earlier work on anomaly detection based
on system calls in that:
\begin{itemize}
    \item A distributed solution for high-volume large-scale
        network is introduced, rather than just an algorithm for
        monitoring of individual processes.
    \item The data representation combines both system-call
        frequencies and temporal dynamics of process behavior.
        The compromise between the amount of information
        preserved and the volume of data collected and processed
        can be tuned continuously with a small set of
        parameters.
    \item An efficient machine learning algorithm based on deep
        architecture, capable of benefiting both from high 
        dimensionality of data and from learning temporal
        features of operating system processes is employed.
\end{itemize}

\section{Methodology Outline and Main Results}
\label{sec:overview}

We approach the following problem: the stream of system calls of
an operating system process is recorded in real time.  Based
on the system call stream, we seek to detect when the behavior
of the process becomes anomalous, either due to misconfiguration
or malfunctioning, or due to malicious activity of an attacker
targeting the process.

An anomalous system call stream may correspond to one or more of
the following scenarios:
\begin{itemize}
    \item \textit{Novelty} --- we cannot classify a process reliably,
        possibly due to malfunctioning or an incompatible
        version.
    \item \textit{Non-grata} --- we identify a process which is known to
       be malicious. 
   \item \textit{Masquerade} --- a process which we reliably classify as
       ‘foo’ presents itself as ‘bar’.
\end{itemize}
Fortunately, all of the above scenarios can be solved through
multiclass classification of processes based on their system
call streams --- and this is indeed the approach we took.
\textit{Novelty} corresponds to classifying a process with low
confidence. \textit{Non-grata} is classifying a process (with
high confidence) as belonging to a known malicious class.
Finally, \textit{Masquerade} relies on the fact that every
running process `presents' itself, that is, sends its own
process name. Masquerade is realized if a process is classified,
with high confidence, to have a different process name than the
name the process pretends to bear. Based on these scenarios,
alerts can be issued and appropriate correcting actions can be
taken.

\subsection{Solution Architecture}

\begin{figure}
    \centering
    \includegraphics[scale=0.4]{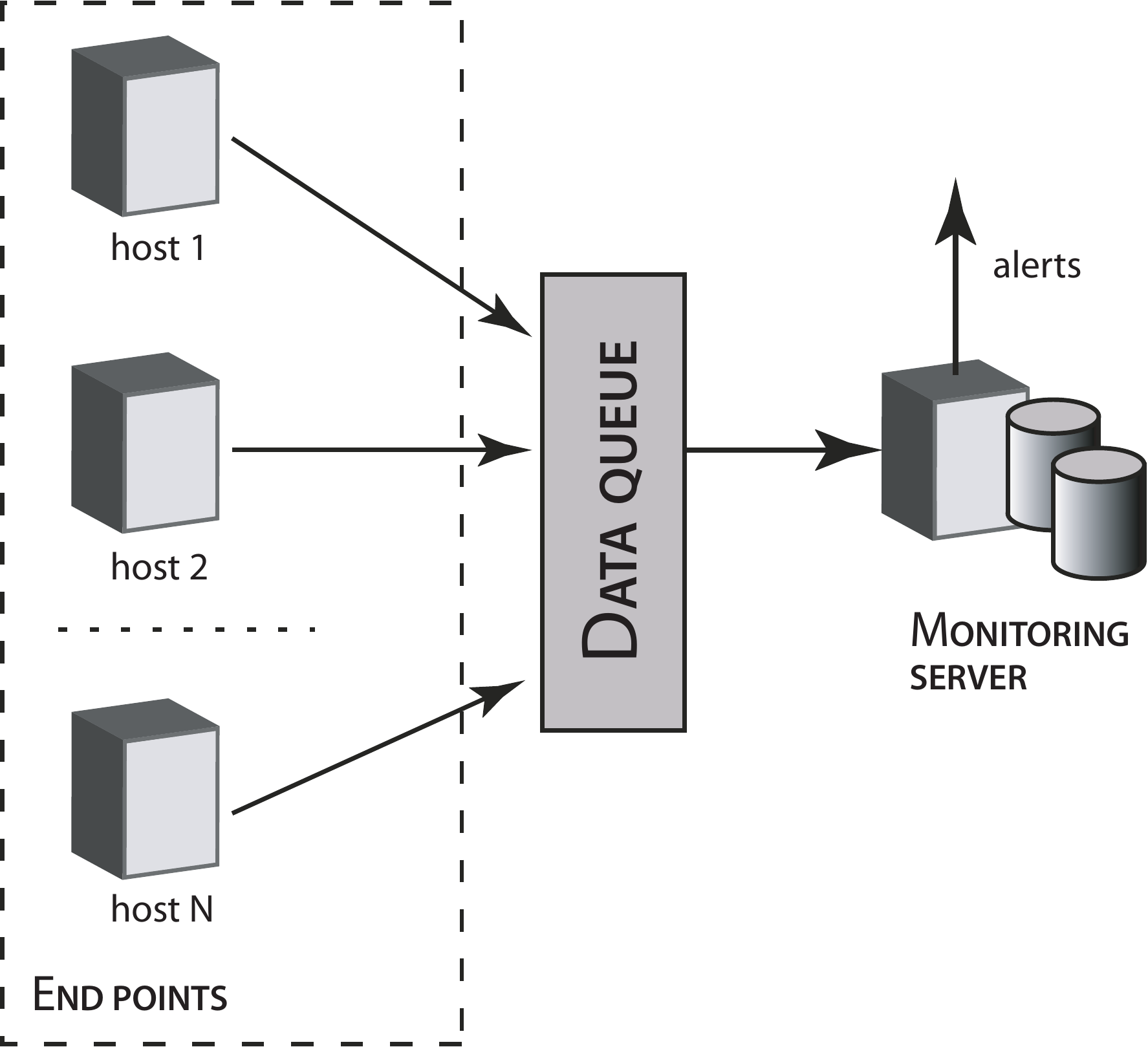}
    \caption{Solution architecture. System call data is sent
    from end points to the monitoring server over the network.}
    \label{fig:architecture}
\end{figure}

The overall architecture of the solution is shown in
Figure~\ref{fig:architecture}. 
The processing is distributed.  The data is collected by an
agent program running on the end points. The data, aggregated
over time frames for efficiency, is sent over the network to the
data queue. The monitoring server consumes the data from the
queue. Based on classification outcomes, the monitoring server
may issue alerts when an anomalous event corresponding to one
of the described scenarios is likely to take place.

There are a few tools~\cite{strace,dtrace,sysdig} for collecting
system calls in real time. Our solution uses
\textit{sysdig}~\cite{sysdig}. Sysdig facilitates efficient
recording of system calls in a high performance environment on
modern Linux systems. 

\subsection{Representation of Data}

The main challenge in implementing the solution is bounding the
amount of data collected and processed while preserving
sufficient information for reliable classification. Using raw
system call logs is infeasible:

\begin{itemize}
    \item A single moderately loaded host can produce a
        million of system  calls per second. Even for a single
        host the task would be challenging. Our architecture
        implies that data from many hosts is sent to the
        monitoring server for centralized processing.
    \item Raw system call logs have long temporal dependencies
        which are hard to learn: two system calls, one relying on
        the outcome of the other, can be hundreds of system calls
        apart.
    \item Although some previous research~\cite{LMH+05,TC06}
        considered using system call arguments for constructing
        features for machine learning, we do not have a compact
        fixed-dimensional representation for system call
        arguments suitable for large-volume training and
        classification.
\end{itemize}

Consequently, we came up with a compact and easily learnable
format based on \textit{sequences of system call count vectors:}

\begin{itemize}
  \item The data is a stream of vectors of integers, each vector
      is $\approx 300$ integers, one per system call type (there are 
       $\approx 300$ system calls in Linux).
   \item Each vector corresponds to a fixed time interval $t$
       (e.g. 1 second).
  \item Each vector component represents the number of calls
      issued during the time interval. 
\end{itemize}

Let us consider an example. In this example we limit the monitoring
to first 6 system calls:
\begin{enumerate}
    \item \texttt{exit}
    \item \texttt{fork}
    \item \texttt{read}
    \item \texttt{write}
    \item \texttt{open}
    \item \texttt{close}
\end{enumerate}
Let us assume that process `foo' performed the following
sequence of system calls during a certain second:

\begin{verbatim}
    fork, open, read, write, read, write,
    read, write, read
\end{verbatim}

The count vector representing the first second is:

\vspace{3pt}
\begin{tabular}{c c c c c c}
    \texttt{exit} & \texttt{fork} & \texttt{read} & \texttt{write} & \texttt{open} & \texttt{close} \\ \hline
    0 & 1 & 3 &  2 & 1 & 0
\end{tabular}
\vspace{6pt}

Then, let us assume that during the next second we observe:

\begin{verbatim}
    write, read, write, close, exit
\end{verbatim}

The corresponding count vector is

\vspace{3pt}
\begin{tabular}{c c c c c c}
    \texttt{exit} & \texttt{fork} & \texttt{read} & \texttt{write} & \texttt{open} & \texttt{close} \\ \hline
    1 & 0 & 1 &  2 & 0 & 1
\end{tabular}
\vspace{6pt}

For input to machine learning model, the count vectors are
normalized and combined into batches. The normalized two-second
batch hence takes the following form:

\vspace{3pt}
\begin{tabular}{r | l l l l l l}
    \hspace{-2em}sec & \texttt{exit} & \texttt{fork} & \texttt{read} & \texttt{write} & \texttt{open} & \texttt{close} \\ \hline
    1 & 0.0 &  0.142 & 0.428  & 0.285 & 0.142 & 0.0 \\ 
    2 & 0.2 & 0.0 & 0.2 & 0.4 & 0.0 & 0.2
\end{tabular}
\vspace{6pt}

Vectors of counts of system calls are collected and sent for
every monitored process at fixed short time intervals.  However,
the monitoring server processes sequences of system call vectors
over longer time spans. This way, the performance guarantee is
maintained through sending fixed amount of data per time unit
independently of the activity on the host, but the temporal
behavior is at least partially preserved. By varying the vector
and sequence time durations, a balance between network and CPU
load, on the one hand, and monitoring accuracy, on the other hand,
can be adjusted depending on performance and accuracy
requirements. 

\subsection{Machine Learning Model}

\begin{figure}
    \centering
    \includegraphics[scale=0.325]{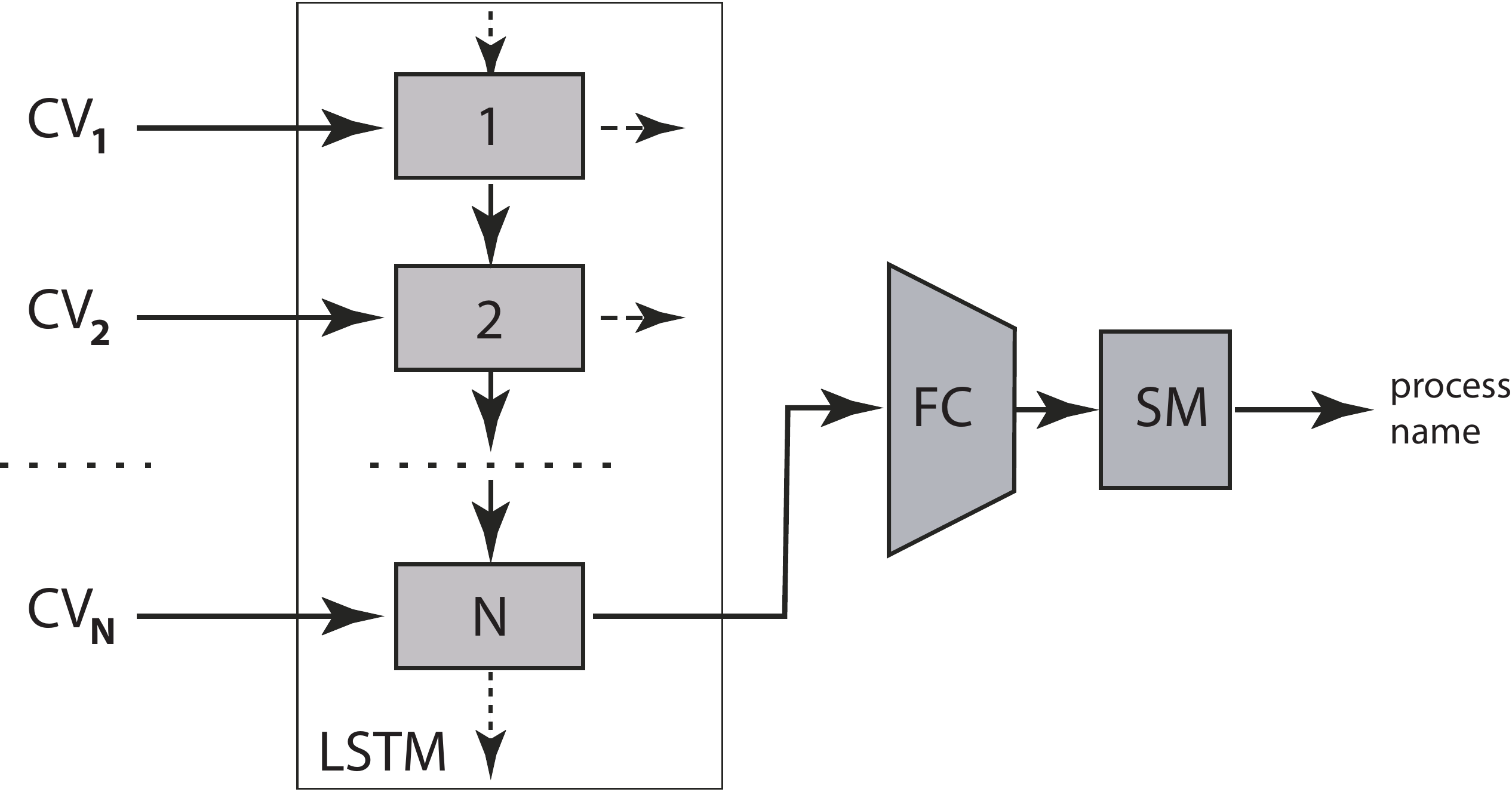}
    \caption{Machine learning model for classification of system
    call streams.}
    \label{fig:model}
\end{figure}
LSTM~\cite{HS97} (Long Short-Term Memory) deep learning
architecture is particularly suitable for processing of
sequences of system call count vectors. We use an LSTM network
to train a model which reliably identifies processes by their
count vector sequences and detects changes in their behavior.

The model structure is shown in Figure~\ref{fig:model}.
The model is composed of a single-layer LSTM, followed by a
fully-connected readout layer (FC), and a softmax layer (SM) for
classification. Details about the machine learning model
and algorithms are provided in Section~\ref{sec:details}.

\subsection{Main Results} 

We evaluated the solution on a laboratory setup and deployed the
solution in the production environment. With 1-second count
vectors and 10-seconds sequence length, the monitoring system
achieves 90-93\% accuracy for all scenarios.  A single
multi-core monitoring server is able to handle a network of
20,000 hosts. Empirical evaluation on the laboratory setup and
the production environment are further described in
Section~\ref{sec:empirical}.

\section{Machine Learning Architecture and Methodology}
\label{sec:details}

System calls are essentially sequential data and preserving the
chronological information is important. Indeed, a sequence of
system calls can be thought of as a sequence of words composing
a sentence, the ordering of the words being critical to identify
the meaning of the sentence. In order to preserve the temporal
aspect of the system calls, we employ an LSTM-based
architecture. LSTM is a type of recurrent neural network
introduced in~\cite{HS97} that maps sequences of variable
lengths to fixed dimensional vectors. It is particularly
suitable for handling sequences of words or systems calls since
the sequences can vary in length. LSTM is quite popular in the
natural language community where it has been successfully 
applied to a vast variety of problems such as speech recognition
or machine translation.

We now describe different variants of architectures that we
experimented with. The architecture depicted in
Figure~\ref{fig:model} represents a neural network composed of a
single-layer LSTM followed by a fully connected layer. The LSTM
receives as input the sequence of count vectors in chronological
order. We refer to this architecture as \textit{simple net}.

\begin{figure}
    \centering
    \includegraphics[scale=0.325]{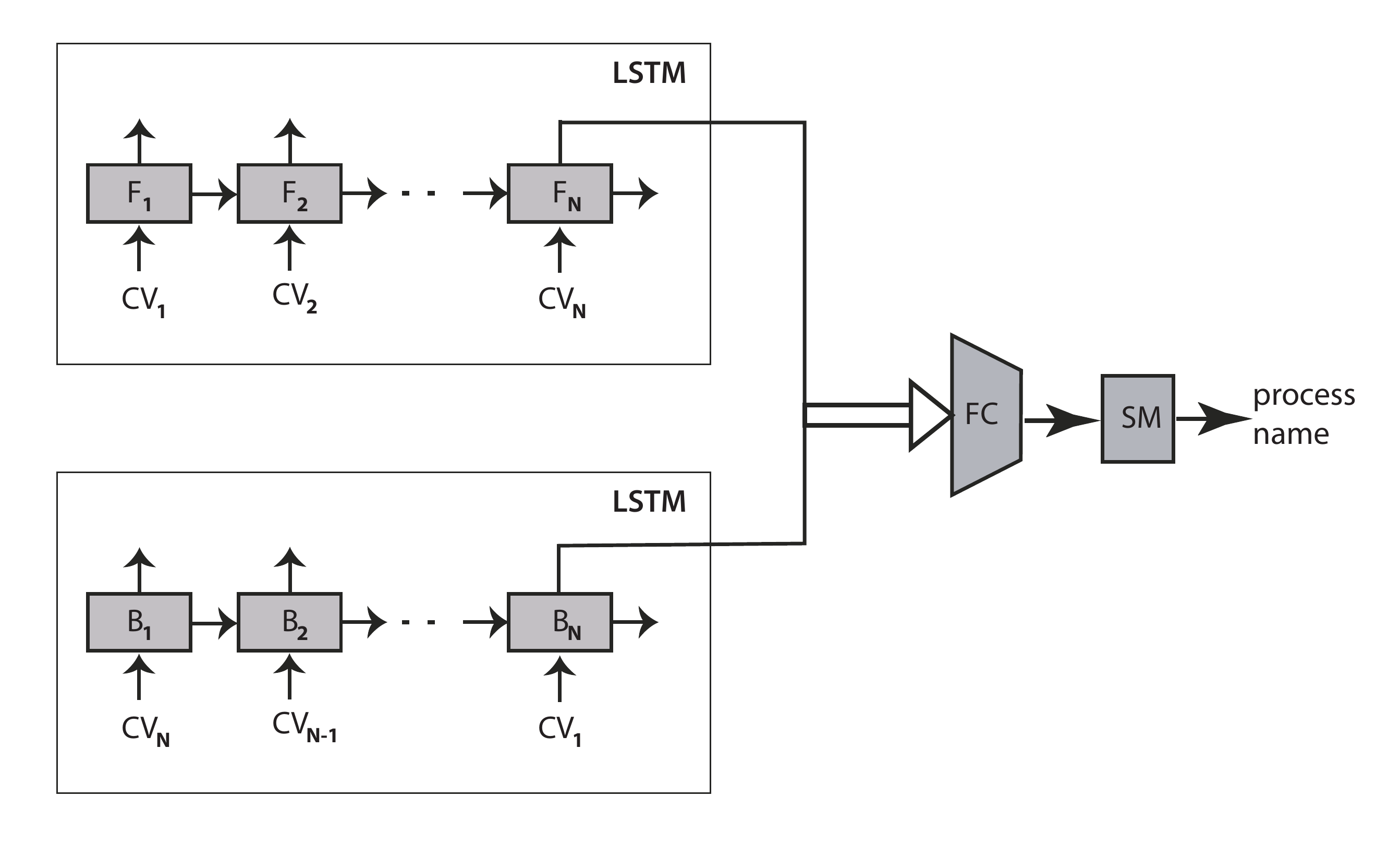}
    \caption{Machine learning model based on bidirectional LSTM.}
    \label{fig:model-bidi}
\end{figure}
A slightly more complex architecture consists in two independent
LSTMs where one receives the sequence in chronological order
while the other receives it in reverse
order (Figure~\ref{fig:model-bidi}). Such a network is
called a bidirectional LSTM. This bidirectional LSTM outputs two
fixed size vectors that are concatenated or averaged and fed to
the following fully connected layer. We refer to this
architecture as \textit{bidirectional net}. A regular LSTM sees
the sequence in chronological order and disregards the
dependence that a later element in the sequence might have on
one that precedes it. By allowing the network to look at the
sequence in reverse order we take into account the dependence of
the sequence in the other direction.

Finally, we experimented with an architecture that we called
\textit{inception-like net} inspired by the inception module
introduced in~\cite{szegedy2015going}. Intuitively, considering
a sequence at multiple scales at the same time, i.e. with
multiple values of the time interval $t$, might give additional
insights. If we take the example of a sentence, considering it
as a sequence of words, but also as a sequence of couple of
words might be useful to better understand the sentence. For an
image, as discussed in~\cite{szegedy2015going}, looking at an
image with sliding windows at various scales helps making sense
of features at different scales. Following this idea, the
\textit{inception-like net} consists in multiple LSTMs with
tied-weights where each of them takes as input the same sequence
but with different values of the time interval $t$. The
different copies of the LSTM output fixed size vectors that are
concatenated and fed to a fully connected layer.

The \textit{simple net} performance is at par or slightly
worse than with the more complex \textit{bidirectional net} and
\textit{inception-like net}. Since the increase in performance
is not significant, we opted for the simplest network. However,
for the sake of completeness, results of the different
architectures are reported in the next section.

\section{Empirical Evaluation}
\label{sec:empirical}

We evaluated our methodology in a laboratory setup as well as in
production environment.\footnote{The code and laboratory data used for the experiments are available at \url{http://github.com/michael135/count-vector-paper-experiments}}

\subsection{Laboratory Setup}

\begin{figure}
    \centering
    \includegraphics[scale=0.325]{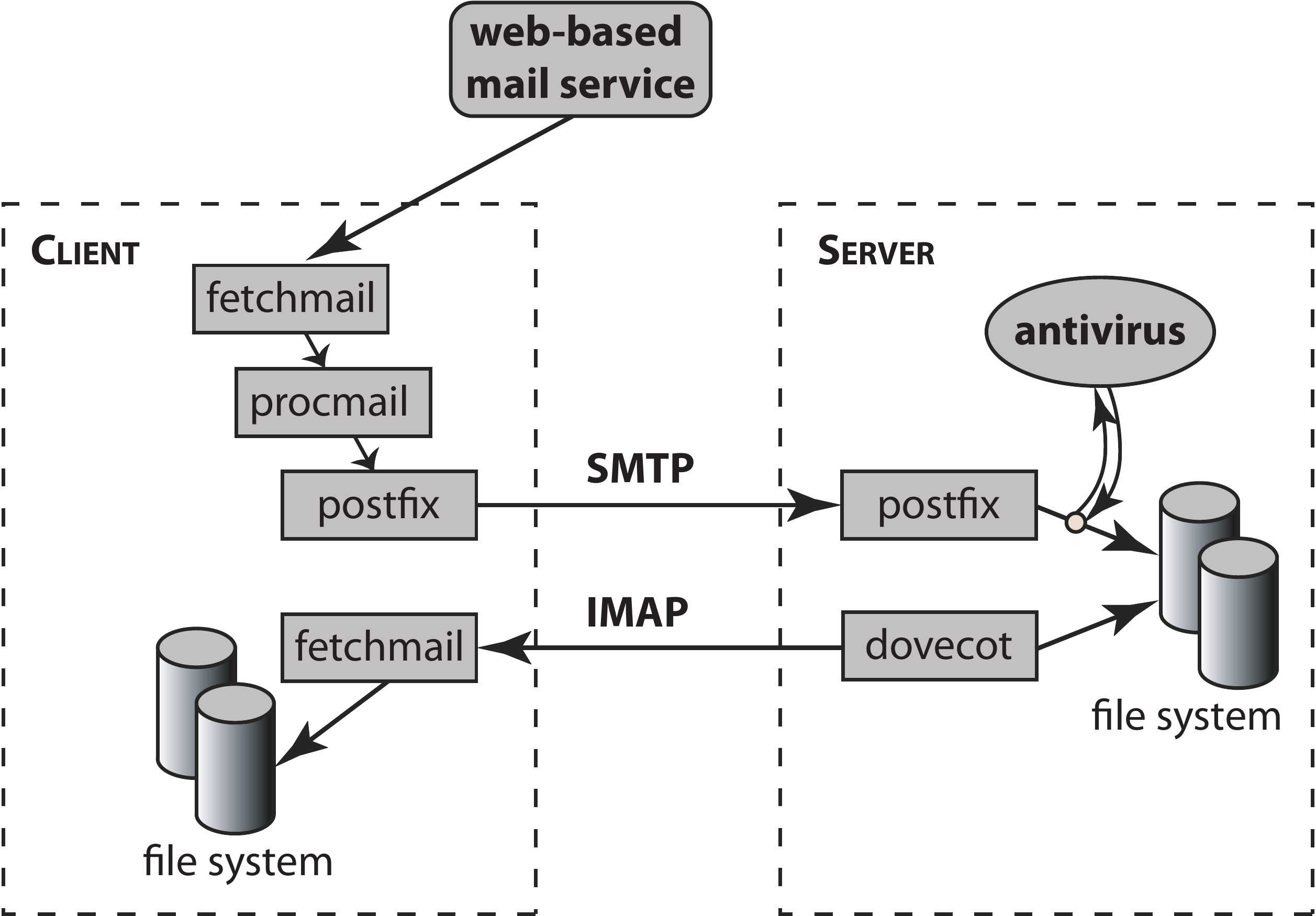}
    \caption{Mail delivery client-server framework for
    evaluating the methodology}
    \label{fig:mailservice}
\end{figure}

For the laboratory setup, we created a data collecting framework
as shown in Figure~\ref{fig:mailservice}.

The setup consists of two hosts: the client and the server. A
number of processes are involved in the workings of the setup.
In the following description, the words in italic correspond to
processes or process groups. The hosts collect emails from an
external server. On the client, \textit{fetchmail} is used to
retrieve emails from a web-based email provider via the IMAP
protocol. Then, \textit{procmail} dispatches received emails,
which are then sent by \textit{postfix} to the server via SMTP
protocol. The server's \textit{postfix} process receives the
emails, passes them through the \textit{amavis} antivirus and
stores in the local filesystem. The \textit{dovecot} process
serves emails via the IMAP protocol. The emails are retrieved by
the client's \textit{fetchmail}, and stored in the filesystem.

\begin{figure}
    \centering
    \includegraphics[scale=0.285]{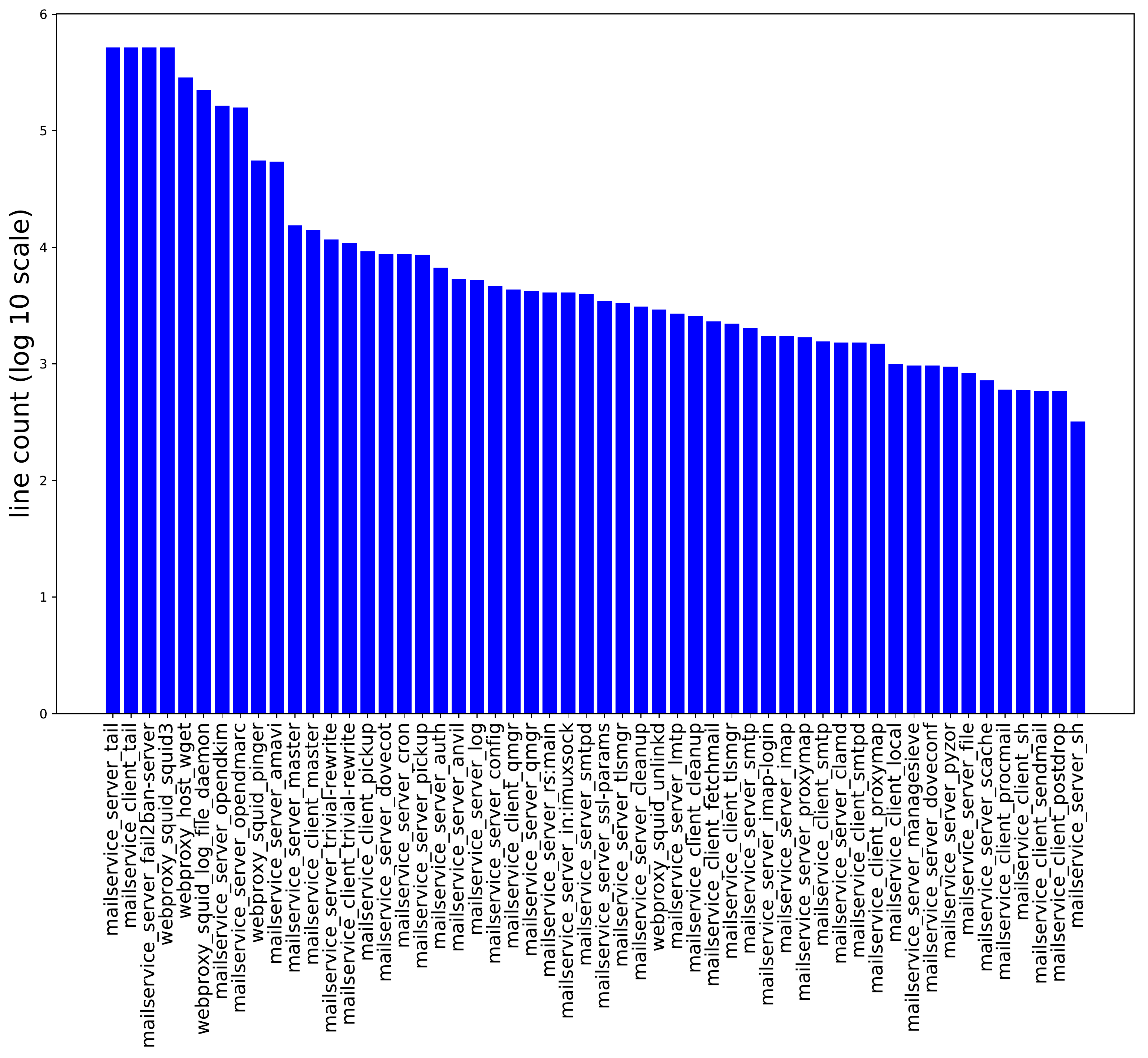}
    \caption{Histogram of count vector counts per process}
    \label{fig:hist_plot}
\end{figure}

In addition to the mentioned processes or process groups, other
utility processes run the hosts and are monitored. The setup is
implemented as Docker~\cite{Docker} containers. In order to
provide sufficient volume and diversity of the data processed,
we opened a dedicated email account with a web-based email
service, and subscribed the email account to many promotion and
notification mailing lists, resulting in the average frequency of
one incoming email per minute. For the empirical evaluation, we
collected data from processes running both on the client and on
the server during one week. The distribution of the count vectors lines is shown Figure~\ref{fig:hist_plot}

For the empirical evaluation, we selected the 28 processes that are
the most represented in our data and trained different models
that aimed at classifying the processes based on their sequence
of system calls. 80\% of the data was used to train the
classifiers and results were calculated on the 20\% left out for
testing. All the results reported for the LSTM-based
architectures used LSTM with 64 hidden units and were trained
using the Adam optimizer with an initial learning rate of
$0.001$. The simple and bidirectional nets used a time interval
$t=1$ while the inception-like net used $t=1,2,3$
simultaneously. L2 regularization was used on the
parameters of the fully connected layer. Dropout on the input
sequence didn't seem to help significantly reduce overfitting in
our experiments so results were omitted.

Results reported with Linear Support Vector Machine
(SVM)~\cite{FCK08}, Logistic Regression~\cite{WD67}, and Random
Forest~\cite{B01} were trained on a single time unit at a time,
i.e. a one-dimensional vector representing $t$ seconds rather
than a block of multiple time units (two-dimensional matrix). At
test time, a process was classified using a majority vote over
the multiple vectors representing it.

All the results reported are in terms of precision and recall
per process (Table~\ref{tab:mailservice}). More precisely, the precisions and recalls over the
different classes are averaged, this metric is known as
macro-average precision and recall.

\begin{table}[h!]
    \centering
    \begin{tabular}{ l | c | c }
        \hline
        Model & Precision & Recall \\
        \hline            
        Logistic Regression & 0.843 (1e-16)  & 0.815 (1e-16) \\
        Linear SVM & 0.850 (1e-16) & 0.827 (1e-16) \\
        Random Forest & 0.860 (0.006) & 0.838 (9e-05) \\
        Simple net & \textbf{0.916} (0.01) & \textbf{0.922} (0.003) \\
        Bidirectional net & \textbf{0.923} (0.01) & \textbf{0.923} (0.003)\\
        Inception-like net & \textbf{0.924} (0.01) & \textbf{0.925} (0.003) \\
        \hline

    \end{tabular}
    \caption{Results for the laboratory setup. Results and
    standard deviations reported in parenthesis were obtained
    with 10 independent runs of the algorithms. A detailed
    description of the models is discussed in the main text.}
    \label{tab:mailservice}
\end{table}

\subsection{Production Environment}

The results in the production environment
(Table~\ref{tab:production}) were obtained by training the
model on data from one set of servers and evaluating the trained
model on the data from a different set of servers, which makes
the task more challenging.  The servers may in general have
different configurations, types and numbers of CPUs, and amounts
of memory installed. Still, the model is able to generalize well
on similar processes among different servers. The experiment was
done on 20 different processes and the hyperparameters of the
models were similar to the ones used for the laboratory setup.

\begin{table}[h!]
    \centering
    \begin{tabular}{ l | c | c }
        \hline
        Model & Precision & Recall \\
        \hline            
        Logistic Regression & 0.791 (1e-16)  & 0.741 (1e-16) \\ 
        Linear SVM & 0.792 (1e-16) & 0.741 (1e-16) \\  
        Random Forest & 0.850 (0.02) & 0.795 (0.02) \\ 
        Simple net & \textbf{0.957} (0.03) & \textbf{0.918} (0.03) \\
        Bidirectional net & \textbf{0.948} (0.04) & \textbf{0.911} (0.04)\\
        Inception-like net & \textbf{0.965} (0.02) & \textbf{0.931} (0.01) \\
        \hline

    \end{tabular}
    \caption{Results for the production environment. Results and
    standard deviations reported in parenthesis were obtained
    with 10 independent runs of the algorithms. A detailed
    description of the models is discussed in the main text.}
    \label{tab:production}
\end{table}

\enlargethispage{-2\baselineskip}
\section{Summary and Future Work}
\label{sec:summary}

The stream of system calls is a rich source of information about
a computer system, but exact processing of the stream is
impractical. Through a novel scheme which enables efficient
processing of the stream while preserving properties essential
for security and health monitoring, we are able to address
several monitoring tasks at large scale with more than
satisfactory accuracy.  Future work is concerned with further
advancing machine learning algorithms, as well as with moving
from plain count vectors to a more compact but still as
informative data representation.

\bibliographystyle{IEEEtran}
\bibliography{refs}

\end{document}